\documentstyle[12pt,epsf]{article}
\newcommand{\be}{\begin{equation}}
\newcommand{\ee}{\end{equation}}
\newcommand{\ba}{\begin{eqnarray}}
\newcommand{\ea}{\end{eqnarray}}
\newcommand{\ps}{\not\! p}
\newcommand{\qs}{\not\! q}
\newcommand{\ks}{\not \! k}
\newcommand{\cs}{\not \! c}
\newcommand{\ds}{\not\!\! D}
\begin{document}
\title{Wess-Zumino-Witten and fermion models
in noncommutative space}
\author{
E.\ Moreno\thanks{Investigador CONICET, Argentina.}\, and\, F.A.\
Schaposnik\thanks{Investigador CICBA, Argentina}
\\
~\\ {\normalsize\it Departamento de F\'\i sica, Universidad
Nacional de La Plata}\\ {\normalsize\it C.C. 67, 1900 La Plata,
Argentina}}
\date{\hfill}
\maketitle
\begin{abstract}
We analyze the connection between  Wess-Zumino-Witten and
free fer\-mion models in two-dimensional noncommutative space.
Starting from the computation of the determinant of the Dirac
operator in a gauge field background, we derive the corresponding
bosonization recipe studying, as an example, bosonization of the
$U(N)$ Thirring model. Concerning the properties of the
noncommutative Wess-Zumi\-no-Witten model, we construct an
orbit-preserving transformation that maps the standard commutative
WZW action into the noncommutative one.
\end{abstract}


\bigskip

%

\section{Introduction}
Noncommutative field theories have recently attracted much
attention in connection with the low-energy dynamics of D-branes
in the presence of a background $B$ field \cite{CDS}-\cite{SW}.
Afterwards, the unusual properties that emerged in the analysis of
such field theories made them very attractive by its own right
\cite{MRS}-\cite{SST}.

Concerning two-dimensional noncommutative field theories, both
bosonic and fermionic models have been recently investigated
\cite{MS}-\cite{NOS}. In particular, we have proven in \cite{MS}
that the fermion determinant of the Dirac operator for fermions in
a gauge field background can be written as a Wess-Zumino-Witten
(WZW)
Lagrangian. As in the ordinary commutative case, this result can
be exploited to establish a bosonization recipe for
non-commutative fermion models, as developed in
\cite{MS},\cite{NOS}.

In this paper we pursue the analysis of two-dimensional
noncommutative models by carefully studying the fermion effective
action. Particular care is taken of the fact that, even in the
$U(1)$ case, the noncommutative Dirac operator can be constructed
in different representations which can be thought of as the
fundamental, anti-fundamental and adjoint ones. As we shall see,
this plays an important role in connection with the resulting
bosonized theory. As an example, we discuss the $U(1)$ massless
Thirring model. We also consider the case in which fermions are
taken in the fundamental representation of $U(N)$. Concerning the
properties of the bosonic model, we establish a
mapping connecting the noncommutative WZW action to the
standard one and discuss its relation with the Seiberg-Witten
\cite{SW} mapping for gauge theories.

The plan of the paper is the following: after establishing our
conventions in section 2, we describe in section 3 the evaluation
of the perturbative effective action for $U(1)$ noncommutative
two-dimensional fermions in a gauge field background. We also
discuss the extension of this calculation to the $U(N)$ case.
Then, we present the exact evaluation of the fermion determinant
both in the fundamental and in the adjoint representations
(sections 4 and 5 respectively) In section 6 we present the
bosonization recipe which allows to infer a mapping between
non-commutative and commutative WZW models (Section 7). In section
8 we apply our results to the analysis of the noncommutative
Thirring model and give our conclusions in section 9.

\section{Two-dimensional fermion determinants in non-commutative space}
We shall work in two-dimensional Euclidean space and define the
$*$ product  between functions $\phi(x)$ and $\chi(x)$ in the form
\be
\left.\phi(x)*\chi (x) = \exp\left(\frac{i}{2} \theta_{\mu
\nu}\partial^\mu_x \partial^\nu_y \right) \phi (x)\chi(y)\right
\vert_{x=y}
\label{moyalp1}
\ee
where  $\theta_{\mu \nu}  = \theta \varepsilon_{\mu\nu}$ with
$\theta$ a real constant. Then, the Moyal bracket is defined as
\be
\{\phi(x),\chi(x)\} = \phi(x)*\chi (x)  - \chi(x)*\phi(x) \ee
which implies a  noncommutative  relation for space-time
coordinates $x^\mu$,
\be
\{x^\mu,x^\nu\} = i\theta_{\mu\nu} \ee
In the case of gauge theories, noncommutativity leads to the
definition of  the curvature $F_{\mu \nu}$ in the form
\be
F_{\mu\nu} = \partial_\mu A_\nu - \partial_\nu A_\mu  -ie
\{A_\mu,A_\nu\} \ee
 Gauge transformations are defined
in the form
\be
A_\mu^g(x) = g(x) * A_\mu(x) * g^{-1}(x) + \frac{i}{e} g^{-1}(x)
\partial_\mu g(x)
\label{asin}
\ee
where $g(x)$  is represented by a $*$ exponential,
\be
g(x) = \exp_*(i\lambda(x)) \equiv 1 + i\lambda(x)  - \frac{1}{2}
\lambda(x) * \lambda(x) + \ldots \label{grupo}\ee with $\lambda =
\lambda^a t^a$ taking values in the Lie algebra of $U(N)$.
The covariant derivative  ${\cal D}_\mu[A]$  implementing gauge
transformations takes the form
\be
{\cal D}_\mu \lambda = \partial_\mu \lambda  -ie\{A_\mu,\lambda\}
\ee

Given a fermion field $\psi(x)$,  three alternative infinitesimal
gauge transformations can be considered  \cite{GM}
\begin{eqnarray}
\delta_\epsilon \psi &=& i\epsilon(x)* \psi(x)  \label{f1}  \\
\delta_\epsilon \psi &=& -i \psi(x) *\epsilon(x) \label{f2} \\
\delta_\epsilon\psi  &=& i \{\epsilon(x),\psi(x)\} \label{f3}
\end{eqnarray}
In this respect, we should refer  to fermions in the  fundamental
$f$ (eq.(\ref{f1})), `anti-fundamental' $\bar f$ (eq.(\ref{f2}))
and `adjoint' $ad$ (eq.(\ref{f3})) representations. The
associated covariant derivative are defined accordingly,
\begin{eqnarray}
D^f_\mu[A]\psi(x) &=& \partial_\mu \psi(x)  - i eA_\mu(x)*
\psi(x)\label{d1}\\
D^{\bar f}_\mu[A]\psi(x) &=& \partial_\mu \psi(x)  + i
e\psi(x) A_\mu(x)\label{d2}\\
D^{ad}_\mu[A]\psi(x)  &=&  {\cal D}_\mu
\psi(x)\label{d3}
\end{eqnarray}
Using each one of these three covariant derivatives,  a gauge
invariant Dirac  action  for fermions can be constructed
\be
S[\bar \psi,\psi,A] = \int d^2x  \bar \psi * i \gamma^\mu D_\mu
\psi \label{efa} \ee
The effective action for fermions in a gauge field
background is defined as
\be
\exp\left(-S_{eff}[A]\right) = \int D\bar\psi D\psi \exp(- S[\bar
\psi,\psi,A]) \label{eff} \ee
\section{Perturbative effective action}

Let us start by computing the quadratic part of the effective
action defined by eq.(\ref{eff}) when the Dirac operator is taken
in the adjoint representation, as defined in eq.(\ref{d3}). The
interaction term $S_I$ of the action $S[\bar \psi,\psi,A]$
(eq.(\ref{efa}))   takes the form, in momentum representation,
\be
S_{I}= 2 i e\;\int_{p,\;q}\; {\bar \psi}(p) \gamma^{\mu} \psi(q)\;
A_{\mu}(-p-q) \sin(q\wedge p) \;, \label{per1} \ee
where $\int_p \equiv \int {d^2p}/{(2\pi)^2}$.

The quadratic part $\Gamma^{(2)}$ of the effective action is given
by the diagram represented in the figure,

\vspace{-2.3 cm}
\epsfxsize=6.in
\epsffile[30 500 480 750]{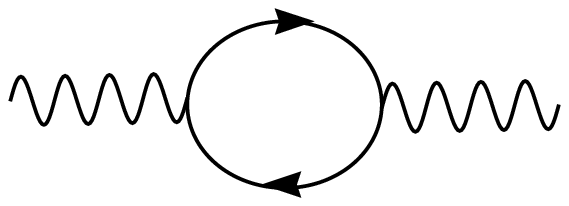}

\vspace{-3.2 cm}

\centerline{Fig. 1}

~

\noindent giving
\be
\Gamma^{(2)} = 2 e \int_{p}\; A_{\mu}(p)\; A_{\nu}(-p) \;
\int_{q}\; \sin(q\wedge p)^2 {\rm tr}\left[\frac{(\qs +
\ps/2)\gamma^{\mu} (\qs - \ps/2) \gamma^{\nu}}{(q+p/2)^2 (q -
p/2)^2}\right]
\label{per2}
\ee

Using the identity $\sin(a)^2 = 1/2 -  \cos(2a)/2$ we can extract
form eq. (\ref{per2}) the so-called planar and non-planar contributions to
$\Gamma^{(2)}$,
\begin{equation}
\Gamma^{(2)} = \int_{p}\; A_{\mu}(p)\; A_{\nu}(-p) \;
\left(\Gamma^{\mu \nu\; (2)}_{planar}  +
\Gamma^{\mu \nu\; (2)}_{nonplanar}
\right)
\end{equation}
with
\begin{eqnarray}
\Gamma^{\mu \nu\; (2)}_{planar} &=& 2 e \int_{q}\;\left\{
\frac{(q^{\mu} + p^{\mu}/2) (q^{\nu} - p^{\nu}/2) + (q^{\mu} -
p^{\mu}/2) (q^{\nu} + p^{\nu}/2)}{(q+p/2)^2 (q - p/2)^2} \right.
\nonumber \\
&& \hspace{1cm} \left. -\frac{\delta^{\mu \nu} (q +p/2) \cdot (q -
p/2)}{(q+p/2)^2 (q - p/2)^2} \right\}
\label{per3}
\end{eqnarray}
\begin{eqnarray}
\Gamma^{\mu \nu\; (2)}_{nonplanar} &=& - 2 e \int_{q}\;\left\{
\frac{(q^{\mu} + p^{\mu}/2) (q^{\nu} - p^{\nu}/2) + (q^{\mu} -
p^{\mu}/2) (q^{\nu} + p^{\nu}/2)}{(q+p/2)^2 (q - p/2)^2} \right.
\nonumber \\
&& \hspace{.5cm} \left. -\frac{\delta^{\mu \nu} (q +p/2) \cdot (q
- p/2)}{(q+p/2)^2 (q - p/2)^2} \right\} \cos(2\; q\wedge p)
\label{per4}
\end{eqnarray}

The planar contribution to the diagram is the standard one and can
be computed using for example dimensional regularization (in this
case the infrared an ultraviolet divergences cancel each other).
One has
\be
\Gamma^{\mu \nu\; (2)}_{planar} = \frac{e}{\pi} \left(\delta^{\mu
\nu} - \frac{p^{\mu} p^{\nu}}{p^2}\right) \label{per5} \ee
It is worthwhile to mention that this result is twice the
effective action in the fundamental and anti-fundamental
representations (that is, taking the Dirac operator as defined
either by (\ref{d1}) or by (\ref{d2})), as it can be easily seen
by noticing that in the later the diagram has a vertex
contribution of $e^{i p\wedge q} \; e^{- i p\wedge q} = 1$ (there
is no non-planar contribution) while in the former we have $ - (2
i \sin(p\wedge q)^2 = 4 (1/2 - 1/2 \cos(2\; p\wedge q))$. Taking
into account that the only the non-oscillating part contributes to
the planar part of the diagram we have
\be
\Gamma^{(2)\; Adj}_{planar} =2\; \Gamma^{(2)\; Fund} \label{per6}
\label{relacion}
\ee
Interestingly enough, this is reminiscent of the relation that one
obtains when one  compares the anomaly and the fermion determinant
for commutative two-dimensional fermions in an $U(N)$ gauge field
background, for the fundamental and the adjoint representation of
$U(N)$. In this case, there is a factor relating the results in
the adjoint and the fundamental which corresponds to the quadratic
Casimir $C(G)$ in the adjoint, as first reported in \cite{PW} (see
for example \cite{FNS} for a detailed derivation). Now, it was
observed in \cite{CG}-\cite{CG2} that diagrams in noncommutative
$U(1)$ gauge theories could be constructed in terms of those in
ordinary non-Abelian gauge theory with $C(G)=2$; this is precisely
what we have found in the present 2-dimensional model. Note that
the explicit form of integrals associated to the diagram in Fig.1
 for each fermion representation can be  constructed if one
take as generators for the Moyal algebra generators
$t$ and $T$ in the adjoint and the fundamental
representation such that
\be
t^q_{pk} = 2 i  \sin(k \wedge p) \delta(k+p-q) \label{gena} \ee
\be
T^q_{pk} = \exp\left(ik\wedge p\right) \delta(k+p-q) \ee

Let us now compute the non-planar contribution to $\Gamma^{(2)}$.
First we exponentiate the propagators with Schwinger parameters
\be
\frac{1}{(q + p/2)^2 (q - p/2)^2} = \int_0^{\infty} dt_1\;
\int_0^{\infty} dt_2\ e^{-t_1 (q + p/2)^2 - t_2 (q - p/2)^2}\;.
\label{per7} \ee

Afterwards we can perform the $q$ integration, which becomes
gaussian, and we get (after the standard redefinition $t_1=u s$
and $t_2= (1-u) s$)) :
\begin{eqnarray}
\Gamma^{\mu \nu\; (2)}_{non-planar} &=& -  \frac{e}{2\pi}
\int_0^{\infty} ds \int_0^1 du\; \left\{\delta^{\mu \nu}\left(
\frac{1}{s} + u(1-u) p^2 + \frac{1}{4s} (\theta\cdot
p)^2\right)\right. \nonumber\\
& & \hspace{-1cm} \left. - 2 p^{\mu} p^{\nu} u (1-u) -
\frac{1}{2s}(\theta\cdot p)^{\mu} (\theta\cdot p)^{\nu}\right\}
e^{- s u(1-u) p^2 - \frac{1}{4s} (\theta\cdot p)^2}\;,
\label{per8}
\end{eqnarray}
where $(\theta\cdot p)^{\mu} = \theta^{\mu \nu} p_{\nu}$. Finally
the $s$ integration can also be performed and we obtain the
expression:
\be
\Gamma^{\mu \nu\; (2)}_{non-planar}= -\frac{1}{2\pi}
\left\{\left(\delta^{\mu \nu} - \frac{p^{\mu} p^{\nu}}{p^2}\right)
- \frac{(\theta\cdot p)^{\mu} (\theta\cdot p)^{\nu}}{(\theta\cdot
p)}\right\}\; \int_0^1 du\;  x K_1(x) \label{per9} \ee
where $x^2=u(1-u) p^2 (\theta\cdot p)^2$ and $K_1(x)$ is the
modified Bessel function of second kind. Notice that the $u$
integral converges.

Now, in two dimensions $\theta^{\mu \nu} = \theta\; \epsilon^{\mu
\nu}$ so we have the identity:
\be
 \frac{(\theta\cdot p)^{\mu} (\theta\cdot p)^{\nu}}{(\theta\cdot
p)^2} = \delta^{\mu \nu} - \frac{p^{\mu} p^{\nu}}{p^2}
\label{per10} \ee
and thus $\Gamma^{(2)}_{non-planar}$ vanishes. That is, up to
quadratic order in the fields, the effective action is given by
the planar part:
\be
\Gamma^{(2)} = \frac{e}{\pi} \int_p A_{\mu}(p) \left(\delta^{\mu
\nu} - \frac{p^{\mu} p^{\nu}}{p^2}\right) A_{\nu}(-p)\; .
\label{per11} \ee

Therefore, assuming that the higher point contributions to the
effective action are the minimal necessary to recover gauge
invariance (we will prove this statement in the next section), the
effective action in the adjoint representation is twice the
effective action in the fundamental representation. Hence, a
relation like (\ref{relacion}) should hold for the complete
effective action $\Gamma$
\be
\Gamma^{Adj} =2\; \Gamma^{Fund} \label{per66}
\label{relacioncom}
\ee

This computation can be generalized to the $U(N)$ case, namely
when the Dirac operator acts as
\be
{\cal D}_\mu \lambda = \partial_\mu \lambda  -ie\{A_\mu,\lambda\}
\ee
with $\lambda\equiv \lambda^a t^a$. In this case the interaction
part of the action takes the form
\be
S_{I}= e \sqrt{2} \;\int_{p,\;q}\; {\bar \psi}^a(p) \gamma^{\mu}
\psi(q)^b\; A_{\mu}^c(-p-q)\left( i d^{abc} \sin(q\wedge p) + i
f^{abc} \cos(q\wedge p) \right) \;, \label{per12}
\ee
where the $f^{abc}$ and $d^{abc}$ are, respectively,  the
antisymmetric and symmetric structure constants of $U(N)$ (the
$\sqrt{2} e$ factor comes from our normalization of the $U(N)$
generators).

The quadratic part of the effective action is in this case
\begin{eqnarray}
\Gamma^{(2)} = \frac{e}{2} \int_{p}\; A_{\mu}^a(p)\;
A_{\nu}^b(-p) \; \int_{q}\; {\rm tr}\left[\frac{(\qs +
\ps/2)\gamma^{\mu} (\qs - \ps/2) \gamma^{\nu}}{(q+p/2)^2 (q -
p/2)^2}\right] \times\nonumber\\
\left\{ \left(d^{a c d} d^{b c d} + f^{a c d} f^{b c
d}\right) + \left(d^{a c d} d^{b c d} - f^{a c d} f^{b c
d}\right)\cos(2 q\wedge p)\right\}
\label{pert13}
\end{eqnarray}

The planar and non-planar contributions are the same as in the
$U(1)$ case except for the group theoretical prefactors.
Taking into account that the non-planar contribution vanishes and
the identity:
\be
d^{a c d} d^{b c d} + f^{a c d} f^{b c
d} = 2 N\; \delta^{ab}
\ee
we finally have
\be
\Gamma^{(2)} = N\;\frac{e}{\pi} \int_p A_{\mu}^a (p) \left(\delta^{\mu
\nu} - \frac{p^{\mu} p^{\nu}}{p^2}\right) A_{\nu}^a(-p)\; .
\label{per14}
\ee

This result is $2 N$ times the quadratic effective action in the
fundamental representation.

\section{The fermion determinant in the fundamental representation:
the $U(1)$ case}

Here we shall briefly describe the exact calculation of the
effective action for noncommutative $U(1)$ fermions in the
fundamental representation as first presented in ref.\cite{MS} by
integrating the chiral anomaly. Indeed, taking profit that in 2
dimensions a gauge field $A_\mu$ can always be written in the form
\be
\not \!\!A  = -\frac{1}{e}\left( i\!\!\not \!\partial U[\phi,\eta]
\right) * U^{-1}[\phi,\eta] \label{amu} \ee
with
\be
U[\phi,\eta] = \exp_*(\gamma_5 \phi + i \eta)  \, , \label{c} \ee
one can relate the fermion determinant in a gauge field background
$A_\mu$ with that corresponding to $A_\mu = 0$  by making a
decoupling change of variables in the fermion fields. As announced,
we consider
in this section  the case of fermions in the fundamental
representation. Then, the appropriate  change of fermionic
variables is
\be
\psi \to U[\phi,\eta]  * \psi \, , \hspace{2cm} \bar \psi \to \bar
\psi * U^{-1}[\phi,\eta] \label{cam} \ee
One gets \cite{MS}
\be
{\det} ( \not \!\partial - ie\!\! \not \!\! A) = \det \not
\!\partial \exp\left( -2 \int_0^1 dt \frac{dJ^f[t\phi,A]}{dt}
\right) \ee
where $J^f[t\phi,A]$ is the Fujikawa Jacobian associated with a
transformation $U_t$ such that  ($o \leq t \leq1$)
\begin{eqnarray}
U_0 &=& 1 \nonumber
\\
U_1 &=& U[\phi,\eta]
\end{eqnarray}
Let us briefly describe at this  point the
calculation presented in \cite{MS} based in the the evaluation of
the chiral  anomaly. Consider an infinitesimal local chiral
transformation which in the fundamental representation reads
\begin{eqnarray}
\delta_\epsilon^5 \psi &=& i \gamma_5\epsilon(x)* \psi(x)
\label{chi1}
\end{eqnarray}
The chiral anomaly ${\cal A} = {\cal A} ^a t^a$,
\be
\partial_\mu j^{a\mu}_5 = {\cal A}^a[A] \, ,
\ee
\be
j^{a\mu}_5 = \bar \psi \gamma_5 t^a \psi \ee
 can be calculated from the Fujikawa Jacobian $J[\epsilon,A_\mu]$
associated with infinitesimal transformation
(\ref{chi1}),
\be
\log J[\epsilon,A_\mu] = -2 {\rm tr^c}\!\!\int \!\!d^2x  {\cal
A}[A]\epsilon(x) =\left. -2 {\rm tr^c}\!\! \int\!\! d^2x {\rm Tr}
\left(\gamma_5 \epsilon(x)\right)\right\vert_{reg} \label{canof}
\ee
Here  Tr means both a trace for Dirac  (tr) and a functional trace
in the space on which the Dirac operator acts while ${\rm tr}^c$
indicates a trace over the gauge group indices. We indicate with
{\it reg} that some regularization prescription should be adopted
to render finite this trace.  We shall adopt the heat-kernel
regularization, this meaning that
\be
{\cal A}[A] = \lim_{M^2 \to \infty} {\rm Tr}\left( \gamma_5
\exp_*\left(\frac{\gamma^\mu \gamma^\nu D_\mu * D_\nu}{M^2}
\right) \right) \ee
The covariant derivative in the regulator has to be chosen  among
those defined  by eqs.(\ref{d1})-(\ref{d3}) according to the
representation one has chosen for the fermions.  Concerning the
fundamental representation,  the anomaly has been computed
following the standard Fujikawa procedure \cite{Fu} in
\cite{AS},\cite{MS}.
\be
{\cal A}^f[A]  = \frac{e}{4\pi} \varepsilon^{\mu\nu}F_{\mu \nu}
\label{anof} \ee (We indicate with $f$ that the fundamental
representation has been considered).  Analogously, one obtains for
the anti-fundamental representation
\be
{\cal A}^{\bar f} [A]  =-\frac{e}{4\pi} \varepsilon^{\mu\nu}F_{\mu
\nu} \label{anti} \ee

Now, writing $\epsilon = \phi dt$,  we can use the results given
through eqs.(\ref{canof})-(\ref{anof})  to get, for the Jacobian,
\be
\log J^f[t\phi,A] = -\frac{e}{2\pi} {\rm tr^c} \int d^2x
\varepsilon_{\mu\nu}F^t_{\mu\nu} *\phi \label{put} \ee
where
\be
F^t_{\mu\nu} = \partial_\mu A_\nu - \partial_\nu A_\mu
-ie\{A^t_\mu,A_\nu^t\} \ee
\be
\gamma^\mu A_\mu^t   =  -\frac{1}{e}\left( i\!\!\not \!\partial
U_t \right) * U_t^{-1} \ee
This result can be put in a more suggestive way in the light cone
gauge where
\begin{eqnarray}
A_+ &=& 0 \nonumber\\ A_- &=&  g(x) *\partial_- g^{-1}(x)
\nonumber\\ g(x) &=& \exp_*(2\phi)
\label{lcg}
\end{eqnarray}
Indeed, in this gauge one can see that (\ref{put}) becomes
\begin{eqnarray}
\log &&\!\!\!\!\!\!\!\!\!\! \left(\frac{{\det} ( \not \!\partial -
ie\!\! \not \! A)}{{\det} \not \!\partial} \right)=\log
J^f[t\phi,A]  = - \frac{1}{8\pi}{\rm tr^c}\int d^2x \left(
\partial_\mu g^{-1}\right) *\left(\partial_\mu g\right) \nonumber\\
&&+  \frac{i}{12\pi}\varepsilon_{ijk}{\rm tr^c}\int_B d^3y
g^{-1}*(\partial_i g) * g^{-1}*(\partial_j g) *g^{-1}*(\partial_k
g)
\end{eqnarray}
Here we have written $d^3y = d^2xdt$  so that the integral in the
second line runs over a 3-dimensional manifold $B$ which in
compactified Euclidean space can be identified with a ball with
boundary $S^2$. Index $i$ runs from 1 to 3. Concerning the
anti-fundamental representation, the calculation of the fermion
determinant follows identical steps.  Using the expression of the
anomaly given by (\ref{anti}), one computes the determinant which
coincides with that in the fundamental representation (remember
that the anomaly is proportional to the charge $e$ while the
determinant to $e^2$). These were the main results in
ref.\cite{MS}. In the next section, we shall extend them to
 the case in
which the Dirac operator is in the adjoint representation.

\section{The determinant in the adjoint representa\-tion:
the $U(1)$ case}

We showed in Section 3 that the effective action for the Dirac
fermions coupled to an external gauge field in the adjoint
representation is, at quadratic order, twice the effective action
in the fundamental representation. In this section we will find
the exact effective action by computing in exact
form  the fermionic determinant
with the heat-kernel approach.

The action is  given by
\be
S_f= \int d^2x  \bar \psi * i \gamma^\mu\left(\partial_{\mu} \psi
+ i e \{A_{\mu}, \psi\}\right) \ee

Writing the field $A_{\mu}$ as in equation (\ref{amu}) (for
simplicity we ill work in the gauge $\eta=0$) we can make a change
if the fermion variables to decouple the fermions from the gauge
field
\begin{eqnarray}
\psi = e^{\gamma^5 \{\phi, \cdot\}} *\chi = \chi + \gamma^5
\{\phi, \chi\}+ \frac{1}{2} \{\phi,\{\phi,\chi\}\}
+\cdots\nonumber\\
{\bar \psi} = {\bar \chi} * e^{\{\cdot,\phi\} \gamma^5}  = {\bar
\chi}  +  \{{\bar \chi},\phi\} \gamma^5 + \frac{1}{2} \{\{{\bar
\chi},\phi\} ,\phi\} +\cdots \label{adcov}
\end{eqnarray}
Similarly to the previous section, the Fujikawa Jacobian
associated to (\ref{adcov}) can bee written as
\be
{\det} ( \not \!\partial + ie \gamma\cdot [A, \cdot]_*) = \det
\not \!\partial \exp\left( -2 \int_0^! dt
\frac{dJ^{ad}[t\phi,A]}{dt} \right) \ee
where $J^{ad}[t\phi,A]$ is the Fujikawa Jacobian associated with a
transformation for fermions in the adjoint,
\begin{eqnarray}
\psi = e^{\gamma^5 [t \phi, \cdot]_*} *\chi \nonumber\\
{\bar \psi} = {\bar \chi} * e^{[\cdot,t \phi]_* \gamma^5}
\end{eqnarray}
where $t$ is a parameter.

Following the same procedure as before, we find that
\be
\frac{d}{dt} \log(J^{ad}) =\int d^2x\; {\cal A}(x)*\phi(x) \ee
where
\be
{\cal A}(x) = 2 \lim_{M \to \infty} {\rm tr}\; \gamma^5\; \int_k
\exp_*({-i k\cdot x}) \exp_*\left({ \not \!\!
D^{2\;Ad}/M^2}\right) \exp_*({i k\cdot x}) \ee
After a straightforward computation it can be proved the following
identity
\be
\exp_*({-i k\cdot x}) \exp_*\left(\ds_{Ad}^{2}/M^2 \right)\;
\exp_*({i k\cdot x}) = \exp_*\left((\ds_{Ad} + i \ks + i e \cs)^2
/M^2\right) \ee
where
\begin{eqnarray}
c_{\mu}(x;k) &=& - 2 i \int_p\; \exp_* \left({-i p\wedge k}
\sin(p\wedge k)\right) A_{\mu}(p) \exp_*({i p\cdot x})\nonumber\\
&=& A_{\mu}(x-\theta\cdot k) - A_{\mu}(k)
\end{eqnarray}
Finally, expanding the exponent up to order $M^{-2}$, taking the
trace and integrating over $d^2 k$ we have
\be
{\cal A}(x) = -\frac{e}{\pi} \lim_{M\to\infty}\; \int_p\;
\left((1-\exp_*\left(^{-M^2 \theta^2 p^2/4} \right)
\right)\epsilon_{\mu \nu} F_{\mu \nu}(p) \exp_*({i p\cdot x}) \ee
Notice that if we take the $\theta\to 0$ limit before taking the
$M\to \infty$ limit ${\cal A}$ vanishes and the Jacobian is $1$,
so we recover the standard (commutative) result which corresponds
to  a  trivial determinant for the trivial $U(1)$ covariant
derivative. However, the limits $\theta\to 0$ and $M\to \infty$ do
not commute so that if one takes the $M^2 \to \infty$ limit at
fixed $\theta$ one has the $\theta$-independent result
\be
\frac{d}{dt} \log J^{Ad}[t\phi,A] = -\frac{e}{\pi} \int d^2x
\varepsilon_{\mu\nu}F^t_{\mu\nu} *\phi \label{puti} \ee
which is twice the result of the fundamental representation. The
integral in $t$ is identical to the one of the fundamental
representation so we finally have
\begin{eqnarray}
\log &&\!\!\!\!\!\!\!\!\!\! \left(\frac{{\det}\left ( \not
\!\partial - ie\gamma_\mu\{A_\mu,~\} \right)}{{\det} \not
\!\partial} \right) = - \frac{1}{4\pi}{\rm tr^c}\int d^2x \left(
\partial_\mu g^{-1}\right) *\left(\partial_\mu g\right) \nonumber\\
&&+  \frac{i}{6\pi}\varepsilon_{ijk}{\rm tr^c}\int_B d^3y
g^{-1}*(\partial_i g) * g^{-1}*(\partial_j g) *g^{-1}*(\partial_k
g)
\end{eqnarray}
We have thus reobtained relation (\ref{per66}) but now by comparing
the exact answer
for the fermion determinant in both representations.

\section{The bosonization recipe}
Once one has gotten an exact result for the fermion determinant,
one can derive the path-integral bosonization recipe. That is,
a mapping from the a two-dimensional noncommutative
fermionic model with onto an equivalent noncommutative bosonic model.
This implies a precise relation between the fermionic and bosonic Lagrangian,
currents, etc. The basic procedure to obtain this bosonization recipe parallels
that already established in the ordinary commutative case. For
the noncommutative case, it was developed in detail in
\cite{NOS}. Here,
we shall describe just the main steps in the derivation
of this equivalence.

The fermion current associated with
the $U(1)$ gauge invariance of the fermion model is naturally given by
\be
j_\mu(x) = \bar \psi(x) * \gamma_\mu  \psi(x)
\label{cor}
\ee
Its coupling to an external source in order to obtain correlation
functions after differentiation can be done simply in the form
\be
S_{int} = \int d^2x j_\mu(x) s_\mu(x) \label{cou} \ee so that
after differentiation with respect to $s$ one gets, as usual,
v.e.v.'s of current correlation functions. Now, being the integral
in (\ref{cou}) quadratic, one can safely replace the ordinary
product between $j$ and $s$ by a $*$ product,
\be
S_{int} = \int d^2x j_\mu(x) * s_\mu(x)
\label{cous}
\ee
One can then write
the fermion generating functional in the presence of a
source $s_\mu$ in the form,
\be
Z_{fer}[s] = \int D\bar \psi D\psi \exp\left(-\int d^2x \bar\psi
D_\mu^{\bar f}[s]\psi
\right)  = \det \left(iD_\mu^{\bar f}[s]
\right)
\label{ri}
\ee
The occurrence of the Dirac operator in the antifundamental
representation is consistent with definition (\ref{cous}) but one
can of course also define the generating functional in terms of
$D_\mu^{f}$.

Since the fermion determinant in the r.h.s. of eq.(\ref{ri})
is gauge invariant, we can write
\be
Z_{fer}[s] = Z_{fer}[s^U]
\label{U}
\ee
where, as before, we indicate with $s^U$ the gauge transformed of $s$
with a gauge group element $U$ as in eq.(\ref{asin}). One now integrates
over $U$ both sides in (\ref{U}) getting
\be
Z_{fer}[U] ={\cal N}  \int D U  \det \left(iD_\mu^{\bar f}[s^U]\right)
\label{UU}
\ee
At this point, one introduces an auxiliary field $b_\mu$ through the
identity
\be
s_\mu^U = b_\mu
\label{end}
\ee
which in the path-integral (\ref{UU}) can be implemented in the form
\be
Z_{fer}[U] =\int D b_\mu \Delta \delta(b_+ - s_+)
\delta(\varepsilon_{\mu\nu} F_{\mu\nu}[b] - \varepsilon_{\mu\nu} F_{\mu\nu}[s])
\det \left(iD_\mu^{\bar f}[b]\right)
\label{UUU}
\ee
Here $\Delta$ can be seen as the Faddeev-Popov determinant associated
with the  delta function ``fixing the gauge'' $b_+ = s_+$.  The
other delta completes the identification (\ref{end}) (These kind of
representations are discussed in detail in \cite{MLNS}). One now
introduces a Lagrange multiplier $a$ in order to represent
the curvature's delta function. Moreover, one writes
\begin{eqnarray}
s_+ &=& i \tilde s^{-1} * \partial_+ \tilde s\nonumber\\
s_- &=& i  s *\partial_-  s^{-1}\nonumber\\
b_+ &=& i  s*b \,\partial_-  (s*b)^{-1}
\label{tries}
\end{eqnarray}
One also trades the Lie algebra valued scalar Lagrange multiplier $a$
by a gauge group valued $\hat a$ through an analogous relation and then, after
repeated use of the Polyakov-Wiegmann identity ends with
\be
Z_{fer}[s] = \int \! D\hat a Db \exp\left((1+2)W[a*\tilde s*s* b]\right)
\exp\left(W[\tilde s *s] - W[\hat a*\tilde s* s]\right)
\ee
We have explicitly written the factor $(1+2)$
in front of the first WZW Lagrangian to stress that this term arises not
only from the fermion determinant but also from the Jacobian arising
in passing from the $b_\mu$ integration to the $b$ integration, a
Jacobian which corresponds to the determinant of the Dirac operator
{\it in the adjoint}.

Use of the Polyakov-Wiegman identity allows to factorize the integral
over $b$ (that is, the $b$ field completely decouples and can be
integrated out). After some manipulations, one finally ends with
\be
Z_{fer}[s] = \int Da \exp\left(-W[\hat a] + \frac{i}{4\pi}
\int d^2x \left(s_+* \hat a* \partial_-\hat a^{-1} + s_- *
\hat a^{-1}* \partial_+\hat a
\right)\right)
\ee
(We have ignored terms quadratic in the source which just give irrelevant
contact terms when computing current-current correlators).

We have then arrived to the identity
\be
Z_{fer}[s] = Z_{bos}[s]
\ee
where $Z_{bos}$ in the r.h.s. is the generating functional for
a Wess-Zumino-Witten model for a gauge-group valued field $a$ coupled
to bosonic sources in such a way that one can already give the
bosonization recipe for currents
\begin{eqnarray}
\bar\psi *\gamma_+ \psi & &  \!\!\!\! \!\!\longrightarrow
\frac{i}{4\pi} \hat a^{-1} *\partial_+ \hat a\nonumber\\
\bar\psi *\gamma_- \psi & &  \!\!\!\!\!\!
\longrightarrow
\frac{i}{4\pi} \hat a *\partial_- \hat a^{-1}
\end{eqnarray}

We have then proved the equivalence between a two-dimensional
noncommutative free fermion model and the non-commutative
Wess-Zumino-Witten model. Now, on the one hand we know that, being
quadratic, the action for noncommutative free fermions coincides
with that for ordinary (commutative) ones. On the other hand we
know that ordinary free fermions are equivalent to a bosonic
theory with ordinary WZW term. The situation can be represented in
the following figure

~

\begin{center}
\unitlength 0.1in
\begin{picture}(23.60,13.05)(22.90,-17.35)
\put(20.1000,-6.0000){\makebox(0,0)[lb]{$
\vphantom{{\rm \,WZW}[a]}
\int \!d^2x\, \bar\psi\! *\! i
\!\!\not
\! \partial \psi$
}}%
\put(21.4700,-18.2000){\makebox(0,0)[lb]{$\int\! d^2x\, \bar\psi i\!\!\not
\! \partial \psi$}}%
\put(42.9000,-18.1000){\makebox(0,0)[lb]{${\rm \,WZW}[a]$}}%
\put(42.9000,-6.0000){\makebox(0,0)[lb]{${\rm WZW}[\hat a]$}}%
\put(22.7000,-12.6000){\makebox(0,0)[lb]{}}%
\put(46.5000,-12.7000){\makebox(0,0)[lb]{?}}%
\special{pn 8}%
\special{pa 2950 530}%
\special{pa 4230 530}%
\special{fp}%
\special{sh 1}%
\special{pa 4230 530}%
\special{pa 4163 510}%
\special{pa 4177 530}%
\special{pa 4163 550}%
\special{pa 4230 530}%
\special{fp}%
\special{sh 1}%
\special{pa 2950 530}%
\special{pa 3017 510}%
\special{pa 3003 530}%
\special{pa 3017 550}%
\special{pa 2950 530}%
\special{fp}%
\special{pn 8}%
\special{pa 4520 1600}%
\special{pa 4520 700}
\special{fp}%
\special{sh 1}%
\special{pa 4520 700}
\special{pa 4500 766}
\special{pa 4520 753}
\special{pa 4540 766}
\special{pa 4520 700}
\special{fp}
\special{sh 1}%
\special{pa 4465 1620}%
\special{pa 4445 1553}%
\special{pa 4465 1570}%
\special{pa 4485 1553}%
\special{pa 4465 1620}%
\special{fp}%
\special{pn 8}%
\special{ar 3590 1130 394 394  2.7109001 6.2831853}%
\special{pn 8}%
\special{sh 1}%
\special{pa 3232 1300}
\special{pa 3200 1260}
\special{pa 3220 1270}
\special{pa 3240 1242}
\special{pa 3232 1300}
\special{fp}%
%
\special{pn 8}%
\special{pa 2950 1730}%
\special{pa 4230 1730}%
\special{fp}%
\special{sh 1}%
\special{pa 4230 1730}%
\special{pa 4163 1710}%
\special{pa 4177 1730}%
\special{pa 4163 1750}%
\special{pa 4230 1730}%
\special{fp}%
\special{sh 1}%
\special{pa 2950 1730}%
\special{pa 3017 1710}%
\special{pa 3003 1730}%
\special{pa 3017 1750}%
\special{pa 2950 1730}%
\special{fp}%
%
\special{pn 8}%
\special{pa 2600 1610}%
\special{pa 2600 700}%
\special{fp}%
\special{sh 1}%
\special{pa 2600 700}%
\special{pa 2570 766}%
\special{pa 2600 753}%
\special{pa 2621 766}%
\special{pa 2600 700}%
\special{fp}%
\special{sh 1}%
\special{pa 2600 1620}%
\special{pa 2580 1553}%
\special{pa 2600 1570}%
\special{pa 2620 1553}%
\special{pa 2600 1620}%
\special{fp}%
\end{picture}%
\end{center}

\vspace{0.3 cm}

\centerline{~ ~ Fig. 2}

~

So, we were able to pass from the noncommutative WZW theory with
action $WZW[\hat a]$  to the commutative one, $WZW[a]$ by going
counterclockwise through the fermionic equivalent models. One
should then be able to find a mapping $\hat a \rightarrow a$,
analogous to the one introduced in \cite{SW} for non-commutative
gauge theories. In the present case, the mapping should connect
{\it exactly} $WZW[\hat a]$ with $WZW[a]$ filling the cycle in
figure 2. Next section is devoted to the construction of such a
mapping.

\section{Mapping Wess-Zumino-Witten actions:\\
Seiberg-Witten change of variables}

Before study the mapping between the non-commutative and standard WZW
theories, let us mention some properties of the Moyal deformation
in two dimensions.

Equation (\ref{moyalp1}) can be re-written in term of
holomorphic and anti-holomor\-phic coordinates in the form:
\be
\left.\phi(z,{\bar z})*\chi(z,{\bar z}) = \exp\left\{\theta (\partial_z
\partial_{\bar w} -
\partial_{\bar z} \partial_w)\right\} \phi (z,{\bar z})
\chi(w,{\bar w})\right \vert_{w=z}
\label{id1}
\ee

This equation reduces considerably in two particular cases. First,
when one of the functions is holomorphic (anti-holomorphic) and
the other is anti-holomorphic (holomorphic), the deformed product
reads as
\be
\phi(z) * \chi(\bar z) = e^{\theta \partial_z \partial_{\bar z}}
\phi(z)\; \chi(\bar z) \;,
\hspace{1cm}
\phi(\bar z) * \chi(z) = e^{-\theta \partial_z \partial_{\bar z}}
\phi(\bar z)\; \chi(z) \;
\label{id2}
\ee
and the deformation is produced by an overall operation over the
standard (commutative) product with no $w\to z$ limit necessary.

Second, if both functions are holomorphic (anti-holomorphic), the
star product coincides with the regular product
\be
\phi(z)*\chi(z) = \phi(z) \chi(z)\; ,
\hspace{1cm}
\phi({\bar z})*\chi({\bar z}) = \phi({\bar z}) \chi({\bar
z})\; .
\label{id3}
\ee
That means that the holomorphic or anti-holomorphic sectors of a
two-dimensional field theory are unchanged by the deformation of
the product. For example, the holomorphic fermionic current in the
deformed theory takes the form,
\be
\hat {\j}_z =\psi^{\dagger}_R\;* \psi_R\; . 
\ee
And since $\psi_R$ has no ${\bar z}$ dependence on-shell, the
deformed current coincides with the standard one
\be
j_z =\psi^{\dagger}_R\; \psi_R\; . 
\ee
Moreover, since the free actions are identical, any correlation
functions of currents in the standard and the $*$-deformed theory
will be identical.

This last discussion tell us that, since the WZW actions are the
generating actions of fermionic current correlation functions,
both actions (standard and non-commutative) are equivalent. It
remains to see if we can link both actions through a
Seiberg-Witten like mapping.
Let us try that.

Consider a WZW action defined in a non-commutative space with
deformation parameter $\theta$. The action is invariant under
chiral holomorphic and anti-holomorphic transformations
\be
\hat g \rightarrow \bar \Omega({\bar z})\hat g\; \Omega(z)
\ee
so in analogy with the Seiberg-Witten mapping we will look for a
transformation that maps respectively holomorphic and
anti-holomorphic ``orbits" into ``orbits". Of course the analogy
breaks down at some point as this holomorphic and anti-holomorphic
``orbits" are not equivalence classes of physical configurations,
but just symmetries of the action. However we will see that such a
requirement is equivalent, in some sense, to the ``gauge orbits
preserving transformation condition" of Seiberg-Witten.

Thus, we will find a transformation that maps a group-valued field
$\hat g'$ defined in non-commutative space with deformation
parameter $\theta'$ to a group-valued field $\hat g$, with
deformation parameter $\theta$. We demand this transformation to
satisfy the condition
\be
\bar \Omega'(\bar z)*' \hat g'*' \Omega'(z) \to \bar
\Omega(\bar z)* \hat g * \Omega(z)
\label{chiraltraf}
\ee
where the primed quantities are defined in a
$\theta'$-non-commutative space and the non primed quantities
defined in a $\theta$-non-commutative space.
In particular this mapping will preserve the equations of motion:
\be
\hat g' = \alpha' *' \beta' \to \hat g=\alpha * \beta
\ee

The simplest way to achieve this, by examining equation
(\ref{id2}) is defining
\be
\hat g[\theta] = e^{-\theta\partial_z\partial_{\bar z}}\; g[0]
\ee
or, infinitesimally
\be
\frac{d \hat g}{d\theta}  = - \partial_z\partial_{\bar z}\;
\hat g
\label{traf1}
\ee
However, the corresponding transformation for ${\hat g}^{-1}$
is more cumbersome
\be
\frac{d {\hat g}^{-1}}{d\theta}  = \partial_z\partial_{\bar z}\;
{\hat g}^{-1} + 2 \partial_{\bar z}({\hat g}^{-1}*\partial_z \hat
g)*{\hat g}^{-1}
\label{traf2}
\ee
So let us consider a more symmetric transformation, that coincides
on-shell, with (\ref{traf1}) and (\ref{traf2}).
Consider thus
\begin{eqnarray}
\frac{d \hat g}{d\theta} &=& \hat g*\partial_{\bar z}\;
{\hat g}^{-1}*\partial_z \hat g \nonumber \\
\frac{d {\hat g}^{-1}}{d\theta} &=& -\partial_z {\hat
g}^{-1}*\partial_{\bar z} \hat g *{\hat g}^{-1}
\label{trafII}
\end{eqnarray}
These equations satisfy the condition (\ref{chiraltraf}) for
functions $\Omega(z)$ and $\bar \Omega(\bar z)$ independent of
$\theta$. Indeed we have, for example
\begin{eqnarray}
\frac{d (\hat g * \Omega(z))}{d \theta}  &=& \frac{d \hat g}{d
\theta}* \Omega(z) - \partial_{\bar z}\hat g * \partial_z
\Omega(z) \nonumber\\
&=& (\hat g*\Omega(z))*\partial_{\bar z}\;(\hat g *
\Omega(z))^{-1} * \partial_z (\hat g* \Omega(z))
\label{inv1}
\end{eqnarray}
and a similar equation for the anti-holomorphic transformation.

The next step is to see how does the WZW action transforms under
this mapping. First consider the variation of the following
object:
\be
\omega = {\hat g}^{-1}*\delta \hat g
\label{omega}
\ee
where $\delta$ is any variation that does not acts on $\theta$.

After a straightforward computation we find that
\be
\frac{d \omega}{d\theta}  = - \partial_{\bar z} \omega * j_z - j_z
* \partial_{\bar z} \omega
\label{trafom}
\ee
where
\be
j = {\hat g}^{-1}*\partial_z \hat g
\label{j1}
\ee
is the holomorphic current. In particular we have
\be
\frac{d j_z}{d\theta}  = - \partial_{\bar z} j_z * j_z - j_z
* \partial_{\bar z} j_z = -\partial_{\bar z} (j_z^2)
\label{trafj}
\ee
Similarly we can find the variations for
\be
\bar \omega = \delta \hat g * {\hat g}^{-1}
\label{bomega}
\ee
and we get
\begin{eqnarray}
\frac{d \bar \omega}{d\theta}  &=&  \partial_{z} \omega * j_{\bar
z} + j_{\bar z}* \partial_{z} \bar \omega \nonumber\\
\frac{d j_{\bar z}}{d\theta}  &=& \partial_{z} j_{\bar z} * j_{\bar
z} + j_{\bar z}* \partial_{z} j_{\bar z} = \partial_z (j_{\bar z}^2)
\label{trafbar}
\end{eqnarray}
where
\be
j_{\bar z} = \partial_{\bar z} \hat g * {\hat g}^{-1} .
\label{barj}
\ee
Note that on-shell, both $j_z$ and $j_{\bar z}$ are
$\theta$-independent, that is the non-commu\-ta\-tive currents
coincide with the standard ones. This result is expected since the
same happens for their fermionic counterparts.

Now, instead of studying how does the $\theta$-map acts on the WZW
action, it is easy to see how does the mapping acts on the {\sl
variation} of the WZW action with respect to the fields. In fact,
we have
\be
\delta W[\hat g] = \frac{k}{\pi}\int d^2x\; {\rm
tr}\left(\partial_{\bar z}j_z\; \omega\right) \label{varwzw}
\ee
where $j_z$ and $\omega$ are the quantities defines in
eqs.(\ref{omega}) and (\ref{j1}) and there is no $*$-product
between them in eq.(\ref{varwzw}) in virtue of the quadratic
nature of the expression.

Thus, a simple computation show that
\be
\frac{d\; \delta W[\hat g]}{d\theta} = 0
\label{st}
\ee
and we  have a remarkable result: the transformation
(\ref{trafII}), integrated between $0$ and $\theta$ maps the
standard commutative WZW action into the noncommutative WZW
action. That is, we have found a transformation mapping orbits
into orbits such that it keeps the form of the action unchanged
provided one simply performs a $\theta$-deformation. This should be
contrasted with the 4 dimensional noncommutative Yang-Mills case
for which a mapping respecting gauge orbits can be found (the
Seiberg-Witten mapping) but the resulting commutative action is
not the standard Yang-Mills one. However, one can see that the
mapping (\ref{trafII}) is in fact a kind of Seiberg-Witten change
of variables.

Indeed, if we consider the WZW action as the effective action of a
theory of Dirac fermions coupled to gauge fields, as we did in
previous sections, instead of an independent model, we can relate
the group valued field $g$ to gauge potentials. As we showed
in eq.(\ref{lcg}), this relation acquires a very simple form in
the light-cone gauge $A_+=0$ where
\be
A_- =  \hat g(x) *\partial_- {\hat g}^{-1}(x)\; .
\ee
But notice that in this gauge, $A_-$ coincides with $j_{\bar z}$
(eq.(\ref{barj}), so we have from equation (\ref{trafbar})
\be
\delta A_- = \delta \theta \left(\partial_{z} A_- * A_- + A_-*
\partial_{z} A_- \right)
\ee
which is precisely the Seiberg-Witten mapping in the gauge
$A_+=0$.


\section{The $U(1)$ Thirring model}
As an example of the bosonization recipe previously obtained,
we analyze in this section the two-dimensional non-commutative Thirring
model with dynamics governed by the (Euclidean) Lagrangian
\begin{equation}
L = \bar \psi * i\!\not\!\partial \psi - \frac{g^2}{2}j_\mu *
j^\mu
\end{equation}
where the fermion current is defined as
\be
j_\mu = \bar \psi * \gamma_\mu \psi \ee
The partition function for the model is
\be
Z = \int D\bar\psi D \psi \exp\left(-\int d^2x L \right) \label{Z}
\ee Let us
introduce an auxiliary vector field $A_\mu$ to eliminate the
quartic fermion self-interaction. We use the identity
\be
\exp\left(\frac{g^2}{2}\int d^2x j_\mu * j^\mu \right) = \int
DA_\mu \exp \left(-\int d^2x \left( \frac{1}{2g^2}A_\mu * A^\mu -
j_\mu * A_\mu\right) \right) \label{hs} \ee
Being all terms in the exponentials quadratic, the $*$ product in
(\ref{hs}) can be replaced by the normal one so that (\ref{hs}) is
just the usual Hubbard-Stratonovich identity.

With this, the partition function  (\ref{Z}) can be written as
\be
Z = \int D\bar\psi D \psi  D A_\mu \exp\left(-\int d^2x L_{eff}
\right) \ee where  we have defined
\be
L_{eff} = \bar \psi * \gamma^\mu iD_\mu^{\bar f}[A] \psi  +
\frac{1}{2g^2} A_\mu * A^\mu \ee Here the Dirac operator $
D_\mu^{\bar f}[A]$ acts on  the anti-fundamental representation
(as defined in eq.(\ref{d2}) with $e=1$). The fermionic
path-integral in $Z$ can now be performed, leading to
\be
Z = \int D A_\mu \exp\left(-\frac{1}{2g^2} \int d^2x A_\mu * A^\mu
\right) \det \left( \gamma^\mu iD_\mu^{\bar f}[A] \right)
\label{Zz} \ee In order to  use the results obtained
 in the previous sections,  we write, for the vector field light-cone
components,
\begin{eqnarray}
A_+ &=&   g^{-1} (x) *\partial_- g(x)\nonumber\\ A_- &=&  h^{-1}
(x) *\partial_- h(x) \label{cambi}
\end{eqnarray}
with $g(x)$ and $h(x)$ group elements, in this case of the $U(1)$
gauge group, as defined in eq.(\ref{grupo}). Then, the fermion
determinant in (\ref{Zz}) can be identified with the
Wess-Zumino-Witten action for the $U(1)$ gauge group element
$gh^{-1}$. Now, such an identification implies that one has
adopted a gauge invariant regularization prescription for
computing the fermion determinant, as done in the precedent
section. A way to see this is the following: one can evaluate
separately the left-handed and right-handed part of the fermion
determinant just by working in the appropriate light-cone gauge,
\begin{eqnarray}
\log\det \left(\partial_+ + i A_+ \right) &=& W[g] \nonumber
\\
\log\det \left(\partial_- + i A_- \right) &=& W[h^{-1}]
\label{doss}
\end{eqnarray}
Then, one can combine the two results to get the determinant of
the complete Dirac operator. But in doing that, one has to add  a
term of the form $\int d^2x A_+A_-$ with an a priori undetermined
coefficient $\bar a$ since such a term is the one involved in the
derivation of a finite answer from the regularized determinants.
One then has
\be
\log \det \left( \gamma^\mu iD_\mu^{\bar f}[A] \right) = W[g] +
W[h^{-1}]  + \frac{\bar a}{4\pi} \int d^2x g^{-1} * \partial_+ g
*h^{-1} * \partial_- h \label{su} \ee
In the case of a gauge theory, gauge invariance fixes the
arbitrary parameter $\bar a = 1$ \cite{PW}. In the present case,
since $A_\mu$ is
 not a gauge field but an
auxiliary field, $\bar a$ remains in principle undetermined,
 this leading, as we shall
see, to the existence of a family of solutions, a well known
feature already encountered in the ordinary Thirring model.

Let us note at this point that a Polyakov-Wiegmann identity can be
seen to hold in noncommutative space,
\be
W[g*h^{-1}] = W[g] + W[h^{-1}] + \frac{1}{4\pi}\int d^2x g^{-1} *
\partial_+ g *h^{-1} * \partial_- h \label{ident} \ee
Using this identity, eq.(\ref{su}) can be rewritten compactly as
\be
\log \det \left( \gamma^\mu iD_\mu^{\bar f}[A] \right) =
W[g*h^{-1}] + \frac{a}{4\pi} \int d^2x g^{-1} * \partial_+ g
*h^{-1} * \partial_- h \label{su1} \ee
where we have redefined the arbitrary parameter so that  $ a =
\bar a -1$.

Concerning the integration variables, we want to write $Z$  as an
integral over $g$ and $h$ and for this we have to take into
account the  Jacobians arising when (\ref{cambi}) is taken as a
change of the path-integral variables,
\begin{eqnarray}
DA_+ DA_- = J[g,h^{-1}] Dg Dh \label{js}
\end{eqnarray}
One can easily see that  Jacobian $ J[g,h^{-1}]$ coincides with
the determinant of a Dirac operator  in the adjoint
representation. Indeed, writing for $g(\alpha) = \exp_*(i\alpha)$
\be
g(\alpha + \delta \alpha) = g(\alpha)* \exp_* (i\delta \alpha) \ee
 an analogous expression for $h$ one has
\be
J[g,h^{-1}] \equiv   \det \frac{\delta A_+}{\delta \alpha}
 \det \frac{\delta A_-}{\delta \alpha'}
\label{unos} \ee or
\be
J^{ad}[g,h^{-1}] =  N \det \left(\gamma_\mu (\partial_\mu+ -
i\{A_\mu,~\}\right) \label{do} \ee
In passing from (\ref{unos}) to (\ref{do}) we have again assumed
that the product of left-handed and right-handed determinants can
be written in terms of the determinant of the complete Dirac
operator in the adjoint at the cost of taking into account
 an in principle undetermined,
regularization dependent coefficient (here included in $N$).

At this point we can use our  previous result on the connection
between the determinant in the fundamental and in the adjoint to
write
\be
\log J^{ad}[g,h^{-1}] =  2 \left( W[g*h^{-1}] + \frac{a}{4\pi} \int
d^2x g^{-1} * \partial_+ g *h^{-1} * \partial_- h \right)
\label{doxx} \ee
where, for consistence, the same regularization dependent
coefficient $a$ has been used.

Then, using eqs.(\ref{su1}),(\ref{doxx}) we can write for
 $Z$
\begin{equation}
Z =  \int Dg Dh \exp \left( 3 W[g*h^{-1}]
 -\frac{1}{2} (\frac{1}{g^2} - \frac{3a}{2\pi})
\int d^2x g^{-1} * \partial_+ g *h^{-1} * \partial_- h \right)
\label{ins}
\end{equation}
Now, the second term in the argument of the exponential  can be
made to vanish if one choses $a$ such that
\be
\frac{1}{g^2} -  \frac{3a}{2\pi} = 0 \label{chosing} \ee
In that case, trading $g(x)$ for a new variable $u(x)$ such that
$g(x) = u(x)h(x)$, one gets
\be
Z = {\cal N} \int Du  \exp(3 W[u]) \label{sli} \ee (Here, the
trivial $h$ integration has been included in a normalization
${\cal N}$).

\noindent We can see this phenomenon from another point of view:
for any choice of the regularization parameter $a$ there exists a
particular value of the coupling constant, which we call $g^*$,
\be
g^{*2} = \frac{2\pi}{3a}
\ee
such that the Thirring model reduces to a Wess-Zumino-Witten model
with ``level'' $k=-3$.

Now, identity (\ref{ident}) can be used to write, instead of
(\ref{sli})
\begin{eqnarray}
Z =  \int\!\!\! Dg Dh \!\!\!\!\! && \!\!\!\! \exp \!\!\!\left(
{\vphantom{\frac{1}{2g^2}}} 3W[g] + 3W[h^{-1}] \right. \nonumber\\
&& \!\!\!\left. -\frac{1}{2} \left(\frac{1}{g^2} - (3a +
1)\frac{1}{2\pi}\right) \int d^2x g^{-1} * \partial_+ g *h^{-1} *
\partial_- h \right) \label{sli2}
\end{eqnarray}
this meaning that there exist a second critical point $g^{**}$
such that
\be
g^{**2} = \frac{2\pi}{1+3a} \ee
at which $Z$ reduces to the
partition function of two (factorized) Wess-Zumino-Witten models,
\be
Z = \int Dg \exp(3W[g]) \int Dh \exp(3W[h^{-1}]) \label{dosfu}
\ee

\section{Discussion}
From the calculation of the Dirac operator determinant in a gauge
field background, we have established the connection between the
action for $U(N)$ fermions and the WZW action in noncommutative
two-dimensional space. The bosonization  recipe for fermion
currents is just the trivial extension of the standard one with
$*$ products replacing ordinary ones so that correlation functions
of currents in the standard and the $\theta$-deformed theory will
be identical. Being the WZW actions the generating functionals of
fermionic current correlation functions, both actions (standard
and non-commutative) are then equivalent. Moreover,  a
Seiberg-Witten like mapping can be constructed so that in both
cases the WZW action is formally the same with the appropriate
product of fields in each case. All these results stem from one of
our fundamental formul\ae (eq.(\ref{st})),
\[
\frac{d\; \delta W[\hat g]}{d\theta} = 0 \]
This formula shows the
independence of the WZW action on the deformation parameter
$\theta$. From this, one can infer that the existence of a mass
scale ($[\theta^{1/2}] = m$) should not affect the scale invariance of
the deformed WZW action. This is consistent with the result of
\cite{FI} where it was found that the one loop beta function for a
nonlinear sigma model with a Wess-Zumino term in noncommutative
space vanishes at the same point as the ordinary model. In our
context, this manifests through the fact that
the noncommutative WZW model, when derived
from the fermion determinant, gives  an action
at its fixed point. This behavior, which is rather evident
at the fermionic
level (being quadratic, the noncommutative free fermion action
coincides with the standard one), is a non-trivial result at the WZW
bosonic level.

On the fermion side, an interesting result is obtained when one
considers the Dirac operator $\not \! \! D$ in different
representations with respect to the $\theta$ deformation. We have
found a factor of $2$ relating the results for $\log \det \not
\!\! D$ in the adjoint and the fundamental, reminiscent of the
factor $C(G)$ that one obtains when one compares the  determinant
for commutative two-dimensional fermions in an $U(N)$ gauge field
background, for the fundamental and the adjoint representation of
$U(N)$.


\newpage

\underline{Acknowledgements}: E.Moreno is partially supported by
Fundaci\'on Antorchas,. This work is partially supported through
grants by  CONICET (PIP 4330/96) and ANPCYT (PICT 97/2285).

\newpage

\end{document}